\documentclass[prb
,twocolumn
,superscriptaddress
,showpacs
]{revtex4-1}

\usepackage[english]{babel}

\usepackage{graphicx}
\usepackage{dcolumn}
\usepackage{amsmath}
\usepackage{amssymb}
\usepackage{hyperref}

\makeatletter
\def\btt#1{\texttt{\@backslashchar#1}}%
\DeclareRobustCommand\bblash{\btt{\@backslashchar}}%
\makeatother

\begin{document}
\title{Influence of the magnetic material on tunneling magnetoresistance and spin-transfer torque in tunnel junctions:
\textit{Ab initio} studies}
\author{Christian Franz, Michael Czerner, and Christian Heiliger}
 \email{christian.heiliger@physik.uni-giessen.de}
\affiliation{I. Physikalisches Institut, Justus Liebig University, Giessen, Germany}
\date{\today}

\begin{abstract}
The dependence of tunneling magnetoresistance and spin-transfer torque in FeCo/MgO/FeCo tunnel junctions on the Co
concentration and the bias voltage are investigated \textit{ab initio}. We find that the tunneling magnetoresistance
decreases with the Co concentration in contradiction with previous calculations but in agreement with recent experiments.
This dependence is explained from bulk properties of the alloys. By using a realistic description of the disorder in the
alloys we can show that even small amounts of disorder lead to a drastic drop in the tunneling magnetoresistance. This
provides a quantitative explanation of the difference between calculated and measured values.

The spin-transfer torque shows a linear voltage dependence for the in-plane component and a quadratic for the
out-of-plane component for all concentrations at small bias voltages. In particular, the linear slope of the in-plane
torque is independent of the concentration. For high bias voltages the in-plane torque shows a strong nonlinear deviation
from the linear slope for high Co concentrations. This is explained from the same effects which govern the tunneling
magnetoresistance.
\end{abstract}



\maketitle


\section{Introduction}

Tunneling magnetoresistance (TMR)~\cite{moodera1995,miyazaki1995} occurs in junctions consisting of two ferromagnetic layers
separated by an insulator. TMR is the change of resistance of the tunnel junction due to changing the relative
orientation between the two magnetizations of the ferromagnetic leads. A number of applications such as hard-disk-drive
read heads, sensors, and magnetic random access memory (MRAM) exploit the effect of TMR. Typical TMR ratios exceed
several hundred percent in crystalline MgO based tunnel junctions~\cite{yuasa2004,parkin2004,heiliger2006} as predicted
theoretically~\cite{butler2001,mathon2001}.

In order to use a magnetic tunnel junction as a storage element in MRAM an efficient way of writing information is
required, i.e.\ of switching the magnetization in one of the layers (free layer). Driving a current through a tunnel
device can switch the magnetic orientation of a ferromagnetic layer. Thereby, one exploits the effect of spin-transfer
torque (STT) which was predicted by Slonczewski~\cite{slonczewski1989,slonczewski1996} and Berger~\cite{berger1996}. The
current gets spin-polarized in one ferromagnetic layer, tunnels through the barrier, and enters the second ferromagnetic
layer. If the magnetization of the second ferromagnetic layer is not perfectly aligned with the polarization of the
current, even due to thermal fluctuations, the transport electrons start to precess around the exchange field of the
second ferromagnetic layer. This in turn leads to a STT acting on the magnetization of this ferromagnetic layer. If the
current is large enough the magnetization can be reversed. For smaller currents the magnetization oscillates, which can
be used to create a microwave oscillator\cite{ralph2008}.

There is a large interest in understanding the behavior of STT in tunnel junctions because this effect is a promising way
to advance the development of MRAM applications~\cite{diao2007,*sun2008,*katine2008}. The critical current where the
magnetization switches is the crucial quantity for applications. However, to lower the critical current one needs to
understand the basic physics, in particular the bias dependence of the STT and the dependence on material parameters.
Experimental results show different bias dependencies~\cite{sankey2008,kubota2008}. In particular, Kubota \textit{et
al.}~\cite{kubota2008} observe a nonlinear bias dependence of the in-plane STT supported by simple model
calculations~\cite{theodonis2006,xiao2008}. In contrast, Sankey \textit{et al.}~\cite{sankey2008} find a linear
dependence of the in-plane STT supported by \textit{ab initio} calculations~\cite{heiliger2008prl}. Recent experimental
investigations by Wang \textit{et al.}~\cite{wang2009} suggest that these differences in the previous experimental
results arise from the analysis of the resonance functions, in particular the dependence on the magnetic offset angles. A
detailed analysis~\cite{wang2009} leads to interpretations supporting the \textit{ab initio} calculations for both
experiments. The experiments~\cite{sankey2008,wang2009} are done with FeCo alloys for the ferromagnetic layers whereas
previous \textit{ab initio} calculations were performed using pure Fe leads. Therefore, the properties of the
ferromagnetic leads in experiments and \textit{ab initio} theory are different. This makes the
agreement~\cite{heiliger2008prl} appear a bit surprising. Here we show that for small voltages the linear dependence of
the in-plane STT is independent of the composition of the FeCo alloy. However, for large voltages the in-plane STT shows
a strong deviation from this linear dependence for large Co concentrations.

It has been demonstrated that imperfect interface structures between the ferromagnetic layer and the barrier, in
particular FeO at the interface, have a strong influence on transport properties in TMR devices~\cite{heiliger2005}.
Recent model calculations show the influence of the size of the exchange splitting and the band filling in the
ferromagnetic layers on the bias dependence of the STT~\cite{kalitsov2009,khalil2010}. These investigations show that the
bias dependence can be drastically changed using different band parameters, which are bulk properties. One task of this
article is to clarify the relative importance of bulk properties of the ferromagnetic layers and interface effects for
the case of perfect interfaces. In this respect, we find that all dominating effects can be understood from bulk
properties.

Prior calculations investigating the TMR for different lead compositions indicate that high Co concentrations should be
beneficial~\cite{zhang2004prb}. These calculations are performed at zero bias and neglect disorder in the alloy. Recent
experiments however find a decrease of the TMR for large Co concentrations~\cite{lee2007,bonell2012}. We show that even
small amounts of disorder cause a substantial decrease of the TMR at zero bias. This provides a quantitative explanation
for the difference between calculated and measured TMR values. At a large bias voltage we find that the TMR decreases
with the Co concentration, in agreement with experiments. This is explained from bulk properties of the FeCo alloys.

The article is organized in the following way. First, we give a short overview over the investigated structures and the
applied methods, in particular the description of the alloys, in Sec.~\ref{sec_method}. In order to understand the
influence of different effects we then investigate the dependence of the TMR and its bias dependence on the ferromagnetic
material. The results are presented in Sec.~\ref{sec_tmr}. The observed high TMR ratios in FeCo/MgO/FeCo tunnel
junctions are related to the STT in the same structures. The origin of the STT in tunnel junctions is explained in
Sec.~\ref{sec_stt}.

\section{Method} \label{sec_method}
\begin{figure}
\includegraphics[width=0.75 \linewidth]{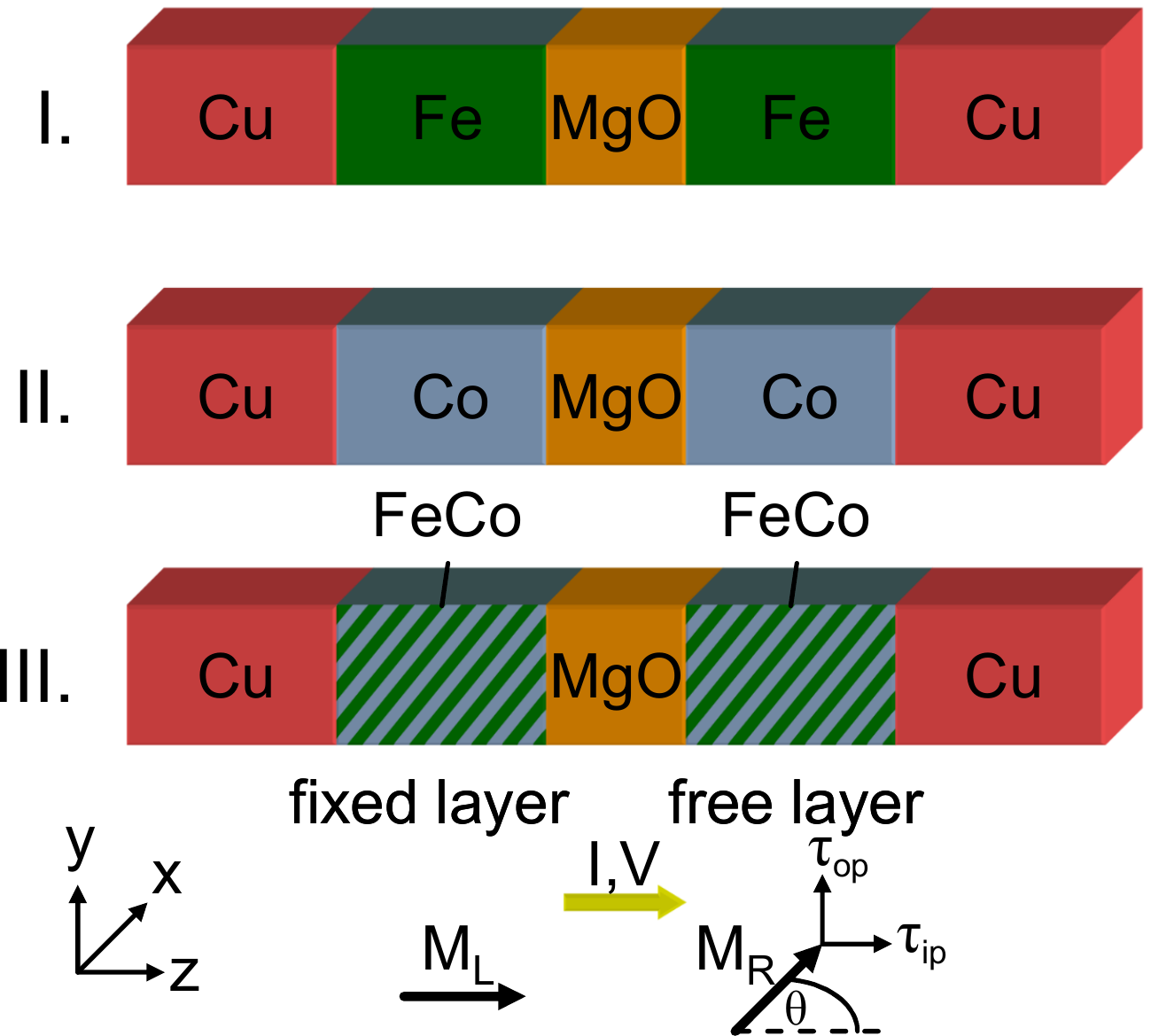}
\caption{(Color online) Structures of the investigated tunnel junctions. In structure III the ferromagnetic layers are
disordered Fe$_{1-x}$Co$_{x}$ alloys. The left $\vec{M}_L$ and right $\vec{M}_R$ magnetizations of the tunnel junctions
lie in the $xz$ plane at a relative angle $\theta$ (here $\theta=90^\circ$). We investigate the spin-transfer torque
acting on the free layer $\vec{M}_R$, while $\vec{M}_L$ is considered fixed. It can be divided into the in-plane torque
$\tau_{ip}$, which lies perpendicular to $\vec{M}_R$ in the plane defined by $\vec{M}_L$ and $\vec{M}_R$, and the
out-of-plane torque $\tau_{op}$, which points perpendicular to that plane. For a positive voltage, the electrons flow
from the free layer to the fixed layer. For junction III we consider either Fe$_{0.5}$Co$_{0.5}$ as an example or the
full concentration dependence.}
\label{fig_sketch}
\end{figure}

We investigate the different junctions shown in Fig.~\ref{fig_sketch}. Each junction consists of 20 monolayers (on
average) FeCo on each side, separated by 6 monolayers MgO. The junction is contacted to artificial copper leads, which
are in Fe-bcc structure. In order to simulate experimental thickness fluctuations, which reduce the effect of quantum
well states, we average over configurations containing 50\%~20 monolayers and 25\% each 19 and 21 monolayers FeCo. The
potentials are calculated self-consistently using a screened Korringa-Kohn-Rostoker (KKR) multiple scattering Green's
function approach and a local-density approximation for the exchange-correlation potential. For the lattice structure we
assume ``ideal" positions: The metals have bcc structure with the equilibrium lattice constant of iron $a_{\mathrm{Fe}}=
0.287\ \mathrm{nm}$. The MgO is strained to $\sqrt{2}\ a_{\mathrm{Fe}}=0.405\ \mathrm{nm}$ in-plane while maintaining
its equilibrium lattice constant $a_{\mathrm{MgO}}= 0.424\ \mathrm{nm}$ out-of-plane. The Fe [100] direction is aligned
with the MgO [110] direction~\cite{butler2001}. The distance between iron and oxygen is $0.235\ \mathrm{nm}$. Note that
this structure differs from the one used in our previous studies, which was based on an experimental structure with FeO
at the interface\cite{meyerheim2001}.

\begin{figure}
\includegraphics[width=0.99 \linewidth]{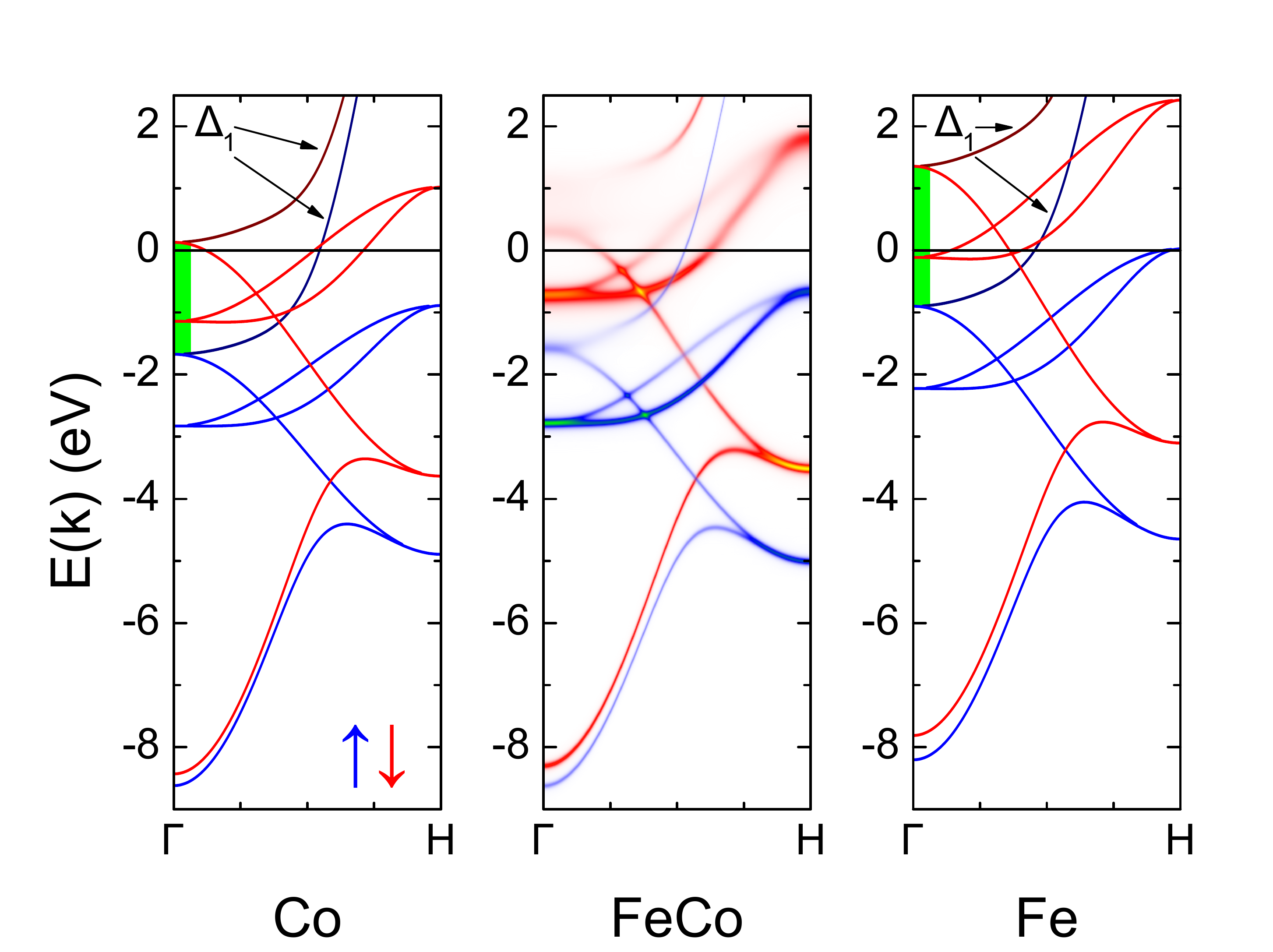}
\caption{(Color online) Band structure of Fe and Co and Bloch spectral function of Fe$_{0.5}$Co$_{0.5}$ along the
$\Delta$-line, i.e.\ $\Gamma$-H. For Fe and Co the $\Delta_1$-half-metallic energy range is marked (range between the
majority ($\uparrow$) and minority ($\downarrow$) $\Delta_1$ band).}
\label{fig_band}
\end{figure}

The alloys are described using the coherent potential approximation (CPA)~\cite{zabloudil}, assuming completely
disordered substitutional alloys. The CPA introduces a complex effective medium which restores the symmetry of the
underlying lattice and accurately describes the scattering of Bloch waves by disorder. This leads to a finite lifetime of
the Bloch states and thus to a broadening of the energy bands. This can be observed in the Bloch spectral
density~\cite{faulkner1980} ($\vec k$-resolved density), see Fig.~\ref{fig_band}. The calculation of transport and
non-equilibrium densities for systems containing CPA-alloys requires determination of non-equilibrium vertex corrections
(NVC)~\cite{franz2013,velicky1969,ke2008}. The NVC describe the influence of the disorder scattering on transport
properties and can be understood as accounting for the diffusive part of the current. The CPA and the NVC have recently
been implemented in our KKR-method~\cite{franz2013}.

The TMR is defined by the ratio
\begin{equation} \label{equ_tmr}
\frac{R_{\mathrm{ap}}-R_{\mathrm{p}}}{R_\mathrm p}= 
\frac{I_{\mathrm{p}}-I_{\mathrm{ap}}}{I_\mathrm{ap}}=
\frac{G_{\mathrm{p}}-G_{\mathrm{ap}}}{G_\mathrm{ap}},
\end{equation}
where $R$ ($I$, ${G\,=\,I/V\,=\,1/R}$) is the resistance (current, conductance) in the tunnel junction for fixed bias
voltage $V$ and parallel (P) or antiparallel (AP) alignment of the magnetizations in the ferromagnetic layers. The
currents are calculated \textit{ab initio} using the non-equilibrium Green's function (NEGF) formalism. Applied to the transport
problem, the NEGF method yields a Landauer formula, which relates the quantum mechanical transmission coefficients $T$ to
the current. Applying a finite bias voltage results in a difference in the chemical potentials $\mu_{L/R}$ in the left
and right lead $V=(\mu_R-\mu_L)/e$ and at zero temperature we get for the current $I_\sigma$ in spin channel $\sigma$
\begin{equation} \label{equ_tr}
I_\sigma=\frac{e}{h} \int_{\mu_L}^{\mu_R} \mathrm d E \sum_{\vec{k}_\parallel} T_\sigma(\vec{k}_\parallel;E),
\end{equation}
where $T_\sigma(\vec{k}_\parallel;E)$ is calculated from NEGFs~\cite{heiliger2008jap,franz2013,henk2006} and
$\vec{k}_\parallel$ is summed over the 2D-Brillouin zone perpendicular to the transport direction. The voltage drop is
assumed to be linear within the barrier. For the limit of zero bias the currents in Eq.~(\ref{equ_tmr}) are replaced by
the corresponding conductances, which are calculated in linear response: $G_\sigma=\frac{e^2}{h} \sum_{\vec{k}_\parallel}
T_\sigma(\vec{k}_\parallel;E_F)$, where $E_F$ is the Fermi-energy. Since our non-relativistic calculations do not include
spin-flip scattering, we get two independent spin channels for collinear magnetizations. Thus, we have
$I_{\mathrm{p}}=I_{\uparrow\uparrow}+I_{\downarrow\downarrow}$ for P and
$I_{\mathrm{ap}}=I_{\uparrow\downarrow}+I_{\downarrow\uparrow}$ for AP alignment, where the double spin indices indicate
the majority($\uparrow$)/minority($\downarrow$) spin in the left and right lead. For alloys we get separate contributions
to the transmission and current accounting for the coherent and diffusive (i.e.\ from the NVC) part~\cite{franz2013}.

The STT consists of two contributions, the in-plane and out-of-plane torque, which are sketched in Fig.~\ref{fig_sketch}.
The in-plane component is zero without an applied voltage whereas the out-of-plane component can be nonzero due to
interlayer exchange coupling~\cite{bruno1994,haney2009}. To calculate the torque $\vec{\tau}_i$ on atomic layer $i$, we
use the change in the magnetic moment in each layer $\delta\vec{m}_i$ due to the current. The torque acting on atomic
layer $i$ is~\cite{haney2007}
\begin{equation} \label{eq_torque}
 \vec{\tau}_i = \frac{d \vec{M}_i}{dt} = \frac{1}{\hbar} \, \Delta_i \, \hat{M}_i \times \delta\vec{m}_i \ , 
\end{equation}
where $\vec{M}_i$ is the magnetic moment, $\hat{M}_i=\frac{\vec{M}_i}{M_i}$, and $\Delta_i$ is the exchange energy on
atomic layer $i$. To obtain the total STT exerted on the free layer $\vec{\tau}_i$ is summed over the corresponding
atomic layers. We use a NEGF technique to calculate the non-equilibrium magnetic moment $\delta\vec{m}_i$. For more
details of our method see Refs.~\onlinecite{franz2013,heiliger2008jap,haney2007}.

In the NEGF calculations we use a $\vec{k}_\parallel$-mesh of $N_k \ge 200^2$ points. For the TMR in the pure limits we
add the requirement that $N_E \, N_k \ge 8 \cdot 10^5$, where $N_E$ is the number of energy points in the integration
(Eq.~(\ref{equ_tr})) to ensure convergence.

\clearpage

\section{Results}

\subsection{Tunneling Magnetoresistance}\label{sec_tmr}

It has been shown that, in order to understand the high TMR in FeCo/MgO/FeCo tunnel junctions, a quantum mechanical
treatment is indispensable. It is a consequence of the symmetry-dependent transmission probability through the MgO
barrier close to the Brillouin zone center and the exchange splitting~\cite{butler2001,heiliger2006}. The states that
dominate the transport properties are of $\Delta_1$ symmetry, i.e.\ states which have the full rotational symmetry of the
interface ($C_{4v}$). In FeCo the exchange splitting leads to an energy gap between the bottom of the majority and
minority $\Delta_1$ band which includes the Fermi-energy. This means that the $\Delta_1$ states, which decay the most
slowly in MgO, are present only for the majority spin in FeCo at the Fermi-level. This $\Delta_1$-half-metallic nature of
FeCo leads to the high TMR ratio.

To be exact, the designation of states in terms of the $\Delta$ representations is only valid at $\overline{\Gamma}$,
yet we will refer to the entire bands with their character at the $\overline{\Gamma}$ point, to simplify notation.

We will show that the major features of the TMR (and also the STT) in the considered junctions with ideal interfaces can
be explained from bulk properties. The importance of the $\Delta_1$ states is a result of the MgO complex band structure,
which determines that these states have the smallest decay rate in the MgO band gap~\cite{butler2001,heiliger2008prb}. We
focus on the properties of the ferromagnetic layers. Figure~\ref{fig_band} shows the band structure of Co and Fe and the
Bloch spectral function of a Fe$_{0.5}$Co$_{0.5}$-alloy along the $\Delta$-line, which coincides with the transport
direction at the $\overline{\Gamma}$ point in the 2D-Brillouin zone. Regarding the pure materials, we see that the change
in band filling caused by one additional electron from $_{26}$Fe to $_{27}$Co leads to a shift of the Fermi-energy, in
particular with respect to the $\Delta_1$-half-metallic region. We find that this has important consequences for the
voltage dependence of TMR and STT. The exchange splitting in Co is smaller than in Fe. As explained in
Sec.~\ref{sec_method}, the broadening in the FeCo Bloch spectral function is a result of the disorder. This obscures the
onset of the $\Delta_1$ band and half-metallic region.

\begin{figure}
\includegraphics[width=0.99 \linewidth]{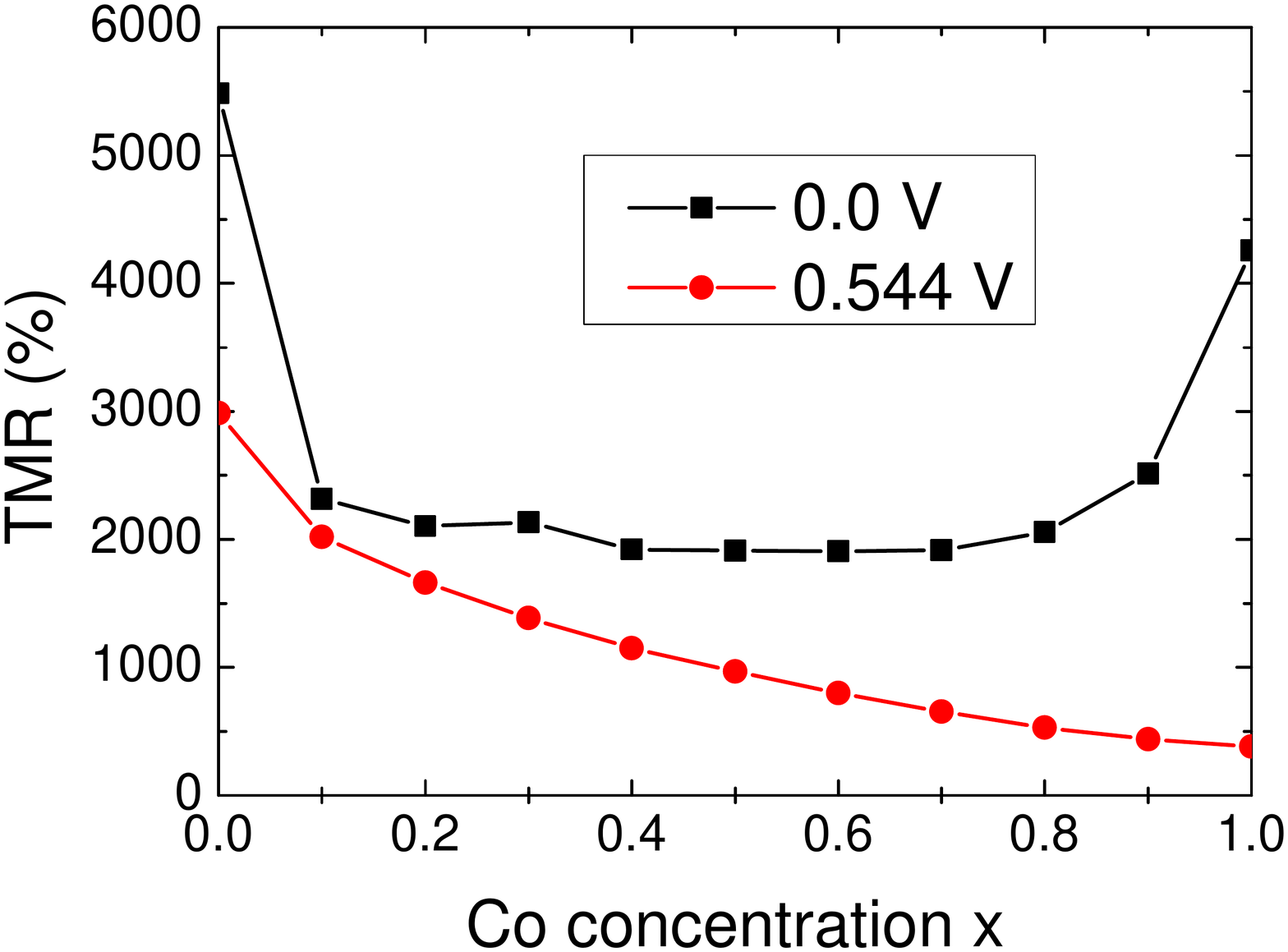}
\caption{(Color online) Concentration dependence of the TMR for zero and a large bias voltage in
Fe$_{1-x}$Co$_{x}$/MgO/Fe$_{1-x}$Co$_{x}$ junctions.}
\label{fig_TMR_c}
\end{figure}

\begin{figure}
\includegraphics[width=0.99 \linewidth]{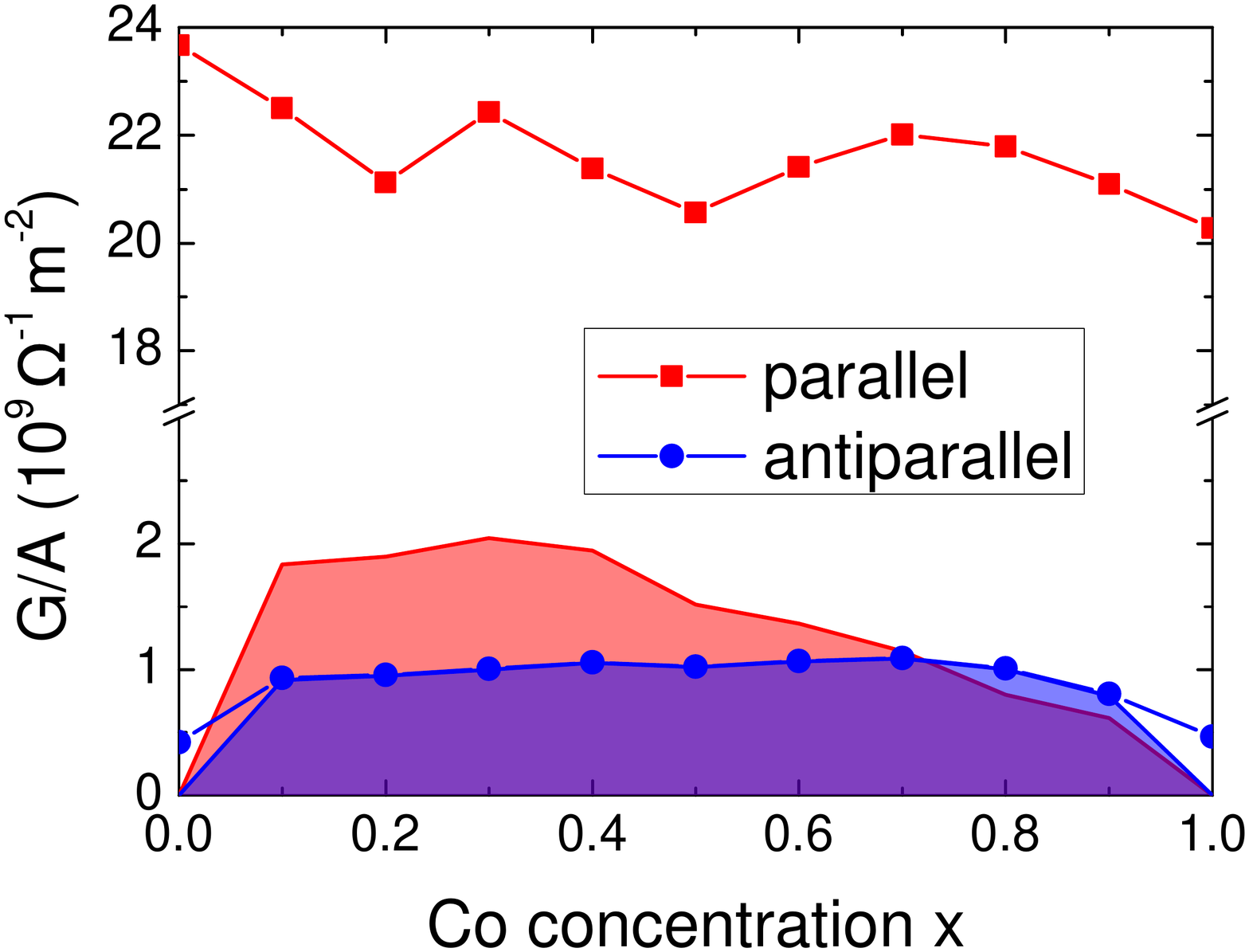}
\caption{(Color online) Concentration dependence of the conductance in Fe$_{1-x}$Co$_{x}$/MgO/Fe$_{1-x}$Co$_{x}$
junctions at zero bias voltage for P and AP alignment. The shaded area indicates the diffusive part of the conductance.
}
\label{fig_trC_V1}
\end{figure}

We start by investigating the concentration dependence of the TMR, calculated using Eq.~(\ref{equ_tmr}). This is shown in
Fig.~\ref{fig_TMR_c}. At zero bias the TMR drops drastically from both pure limits to finite concentrations but then
remains nearly constant throughout the concentration range. At the large voltage the TMR is smaller and decreases with
the Co concentration. The full voltage dependence is discussed later. In order to understand the striking dependence at zero
bias we analyze the dependence of the tunneling conductance in the P and AP configuration shown in Fig.~\ref{fig_trC_V1}.
The drop in the TMR is caused by an increase of the conductance in the AP configuration, while the P conductance remains
roughly constant. From Fig.~\ref{fig_trC_V1} we find that the AP conductance is completely diffusive. This indicates
that the disorder scattering reduces the effects responsible for the high TMR.

\begin{figure}
\includegraphics[width=0.99 \linewidth]{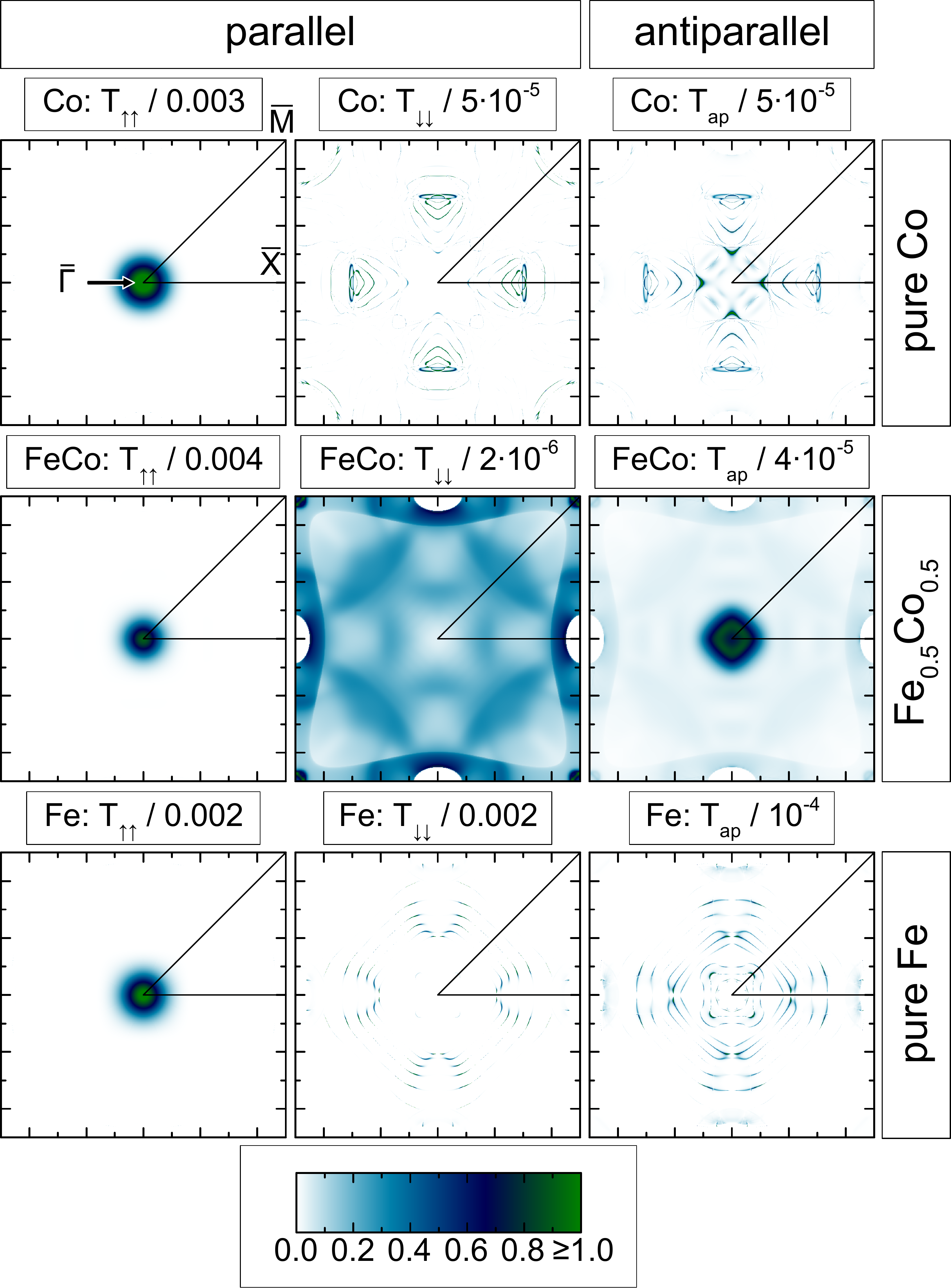}
\caption{(Color online) $\vec{k}_\parallel$-resolved transmission $T_{\sigma\sigma'}(\vec{k}_\parallel;E_F)$ at zero bias
for junctions containing 20 monolayers Co, Fe$_{0.5}$Co$_{0.5}$, or Fe in each ferromagnetic layer (no thickness
averaging) at the respective Fermi-energy for the different spin channels. For the AP alignment we show
$T_{ap}=T_{\uparrow\downarrow}+T_{\downarrow\uparrow}$. Some sharp peaks are clipped to improve the overall visibility.
($\overline{\Gamma}=\frac{2\pi}{a}(0,0)$, $\overline{X}=\frac{2\pi}{a}(1/2,0)$, and $\overline{M}=\frac{2\pi}{a}(1/2,1/2)$)}
\label{fig_tr_k_all}
\end{figure}

To explicate this, we show the $\vec{k}_\parallel$-resolved transmission at zero bias in Fig.~\ref{fig_tr_k_all}. This
allows us to explain the origin of the large TMR in more detail.
Note that for finite concentrations the effective $\vec{k}_\parallel$-resolved transmission can be understood as the
ratio of incident and transmitted electrons at $\vec{k}_\parallel$ within the current, where the incident and
transmitted particles need not be the same.
For all junctions, the P majority channel $T_{\uparrow\uparrow}$ shows a single pronounced peak around
$\overline{\Gamma}$. This peak consists of majority $\Delta_1$ states, which are strongly facilitated by the MgO complex
band structure in this $\vec{k}_\parallel$-region~\cite{heiliger2008prb}. It adds up to a large $G_{\uparrow\uparrow}$
conductance which dominates $G_p$. It is important to note that this peak cannot be explained from the
$\vec{k}_\parallel$-resolved density of states (bulk or interface) but requires knowledge of the character of the states.
For Fe and Co the P minority channel $T_{\downarrow\downarrow}$ shows a complicated structure. The gross shape can be
explained from the MgO complex band structure. Most importantly, it causes the empty spot around $\overline{\Gamma}$,
where the available minority states are suppressed. The details of $T_{\downarrow\downarrow}$ depend on several effects,
including the shape of the Fermi-surface, quantum well states, and interface resonance states. For AP alignment the
majority (minority) states of the left lead tunnel to the minority (majority) states in the right lead. Since the
states contributing to $T_{\uparrow\uparrow}$ and $T_{\downarrow\downarrow}$ are located in different
$\vec{k}_\parallel$-regions they do not overlap. This leads to the strong suppression of the transmission $T_{ap}$ in the
AP alignment. Because of that, we have $G_p \gg G_{ap}$ and thus a high TMR. Note that this requires the full
$\vec{k}_\parallel$-resolved information and cannot be obtained from integrated properties. In particular, this does not
require $G_{\uparrow\uparrow}\gg G_{\downarrow\downarrow}$, i.e.\ a large polarization of the P conductance.

For Fe$_{0.5}$Co$_{0.5}$ we find that $T_{\downarrow\downarrow}$ as well as $T_{ap}$ are strongly smeared out. This is an effect of the
disorder which leads to a scattering of the Bloch waves and thus redistributes the electrons across the Fermi-surface.
The Fermi-surface exhibits the same broadening that is visible in the Bloch spectral density. In $T_{\uparrow\uparrow}$ this
effect is not visible because it is dominated by the coherent contribution. This is expected from the Bloch spectral
function in Fig.~\ref{fig_band}, which shows a very small broadening of the majority $\Delta_1$ band at the Fermi-energy,
indicating a weak scattering and thus a mainly coherent transport. On the other hand, the minority bands show a strong
broadening and are therefore strongly affected by scattering leading to a mostly diffusive transport (compare
Fig.~\ref{fig_trC_V1}). The redistribution increases the overlap between the states contributing in
$T_{\uparrow\uparrow}$ and $T_{\downarrow\downarrow}$ which causes the observed increase in $G_{ap}$ compared to the pure
materials.

Consequently, the striking concentration dependence at zero bias is a result of including disorder in the description. It
cannot be reproduced by approximating the alloy with an ``ordered alloy"~\cite{zhang2004prb}, i.e.\ a stacking of atomic
Fe and Co layers, which has not been observed in experiments. Obviously, omitting the diffusive contributions (i.e.\
neglecting the NVC) does not lead to a meaningful result. We remark that the very high TMR values for the pure components
depend sensitively on the computational details. In particular, they show a strong variation for the different
thicknesses entering in the thickness averaging, e.g.\ between 1380\% and 14300\% for Co. This variation decreases with
the bias voltage. This sensitivity explains some deviations between different values presented in literature. On the
other hand, we know from Fig.~\ref{fig_TMR_c} that small amounts of disorder in the layers or at the interfaces reduce
the TMR severely. This makes it very hard to achieve the theoretical values in experiments.

\begin{figure}
\includegraphics[width=0.99 \linewidth]{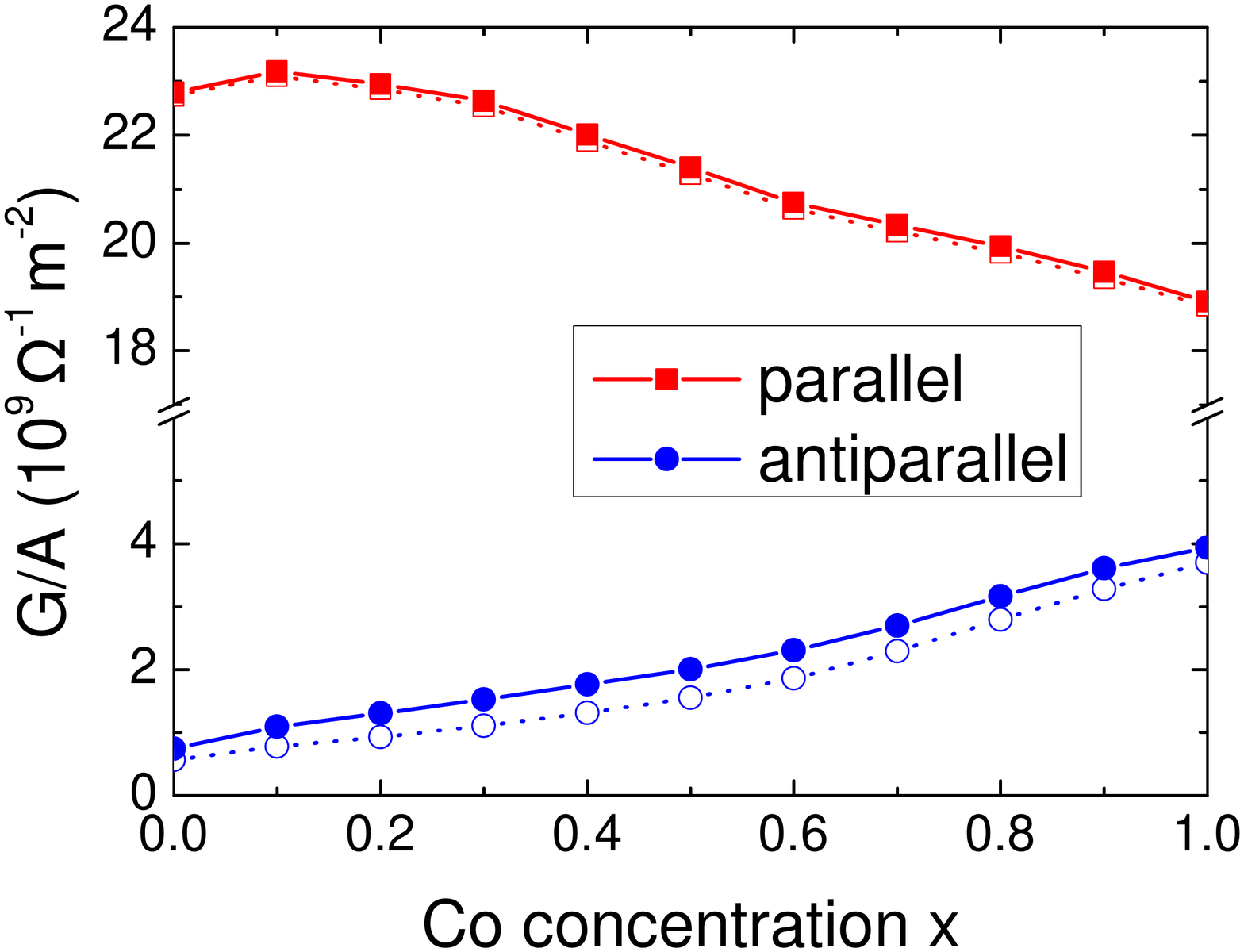}
\caption{(Color online) Concentration dependence of the conductance in Fe$_{1-x}$Co$_{x}$/MgO/Fe$_{1-x}$Co$_{x}$
junctions at a bias voltage $V=0.544 \ \mathrm{V}$ for P and AP alignment. The dashed lines indicate the dominating
contribution, which is the $\uparrow\uparrow$ channel for P and the $\downarrow\uparrow$ channel for AP alignment.}
\label{fig_trC_V2}
\end{figure}

At the large voltage of $0.544 \ \mathrm{V}$ the current in the P configuration decreases slightly while the current in
the AP configuration increases linearly with the Co concentration, leading to a decrease of the TMR. The corresponding
conductances are shown in Fig.~\ref{fig_trC_V2}. The origin of this dependence is very different from that at zero bias.
The concentration dependence of the AP current, which primarily determines that of the TMR, is governed by the
$\downarrow\uparrow$ channel. The origin of this dependence is related to the position of the Fermi-energy relative to
the $\Delta_1$-half-metallic region. This will be explained in more detail later.

Our results clearly favor small Co concentrations in order to obtain a large TMR. The concentration dependence of the TMR
was investigated in recent experiments~\cite{lee2007,bonell2012}. Both experiments find a drop in the TMR for large Co
concentrations in agreement with our prediction. On the other hand, they find a maximum TMR for a Co concentration of
about 25\% and a decrease towards pure Fe. There are no indications for this decrease in our calculations suggesting that
it is caused by an effect, which was not considered. Still, the TMR values in these experiments (up to 500\% at low
temperature) are a factor of 3-4 smaller than the theoretical predictions. This might be related to an imperfect lattice
structure, in particular at the interfaces.

\begin{figure}
\includegraphics[width=0.99 \linewidth]{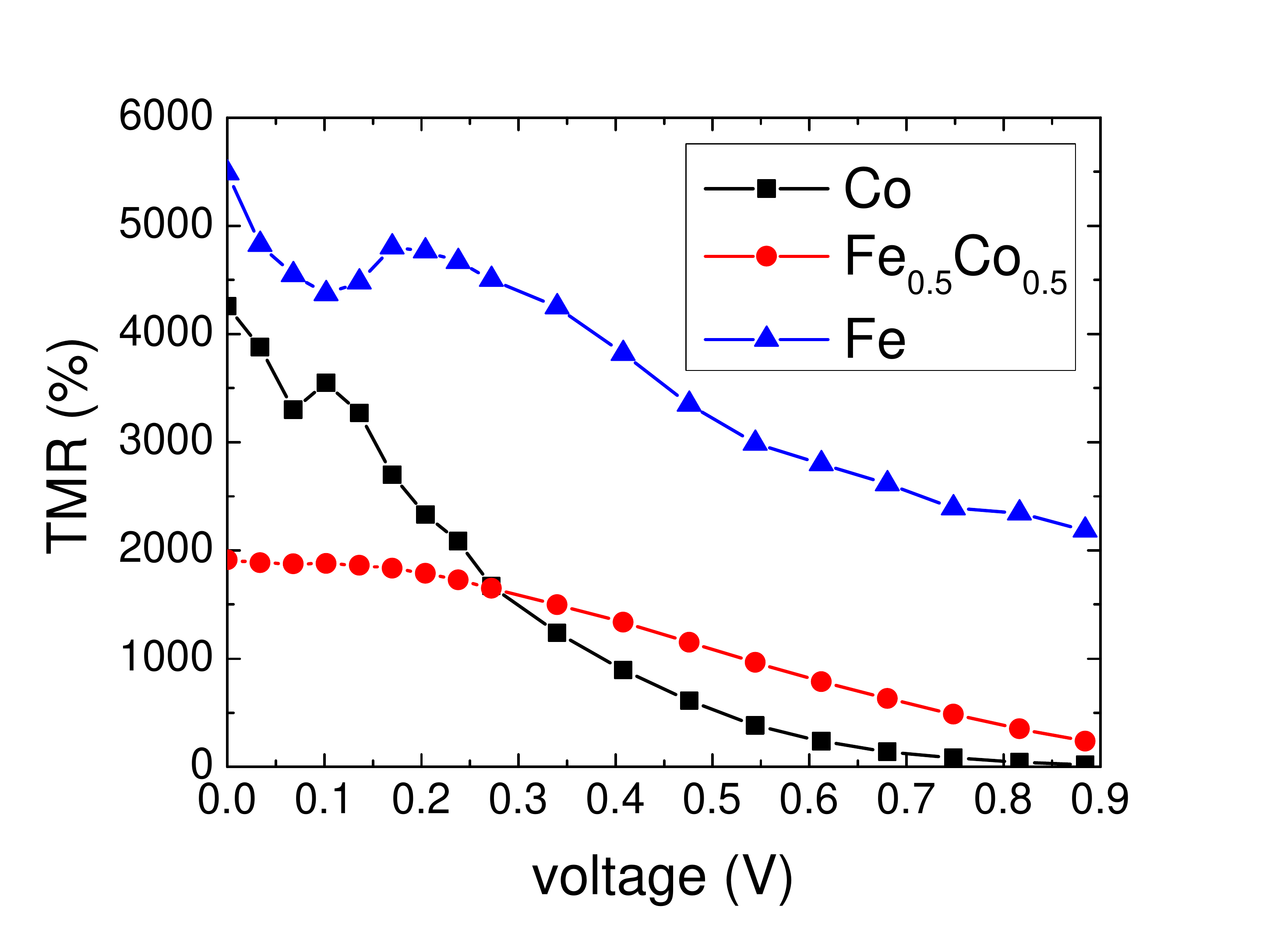}
\caption{(Color online) Bias voltage dependence of the TMR in junctions with Fe, Fe$_{0.5}$Co$_{0.5}$, and Co.}
\label{fig_TMR_V}
\end{figure}

Figure~\ref{fig_TMR_V} shows the voltage dependence of the TMR for the pure materials and for Fe$_{0.5}$Co$_{0.5}$
ferromagnetic layers. Since we consider only symmetric junctions, the voltage dependence is symmetric. The strong
features at low bias for pure Fe and Co can be attributed to contributions from tunneling between quantum well states in
the ferromagnetic layers, which were not completely removed by the thickness averaging process. Both pure cases start at
very high values, but the TMR value for Co leads decreases much faster with increasing voltage. The TMR value for
Fe$_{0.5}$Co$_{0.5}$ leads is much smaller at zero bias, but it decreases slower with the bias voltage than for Co, and
thus eventually becomes larger than the Co value.

\begin{figure}
\includegraphics[width=0.99 \linewidth]{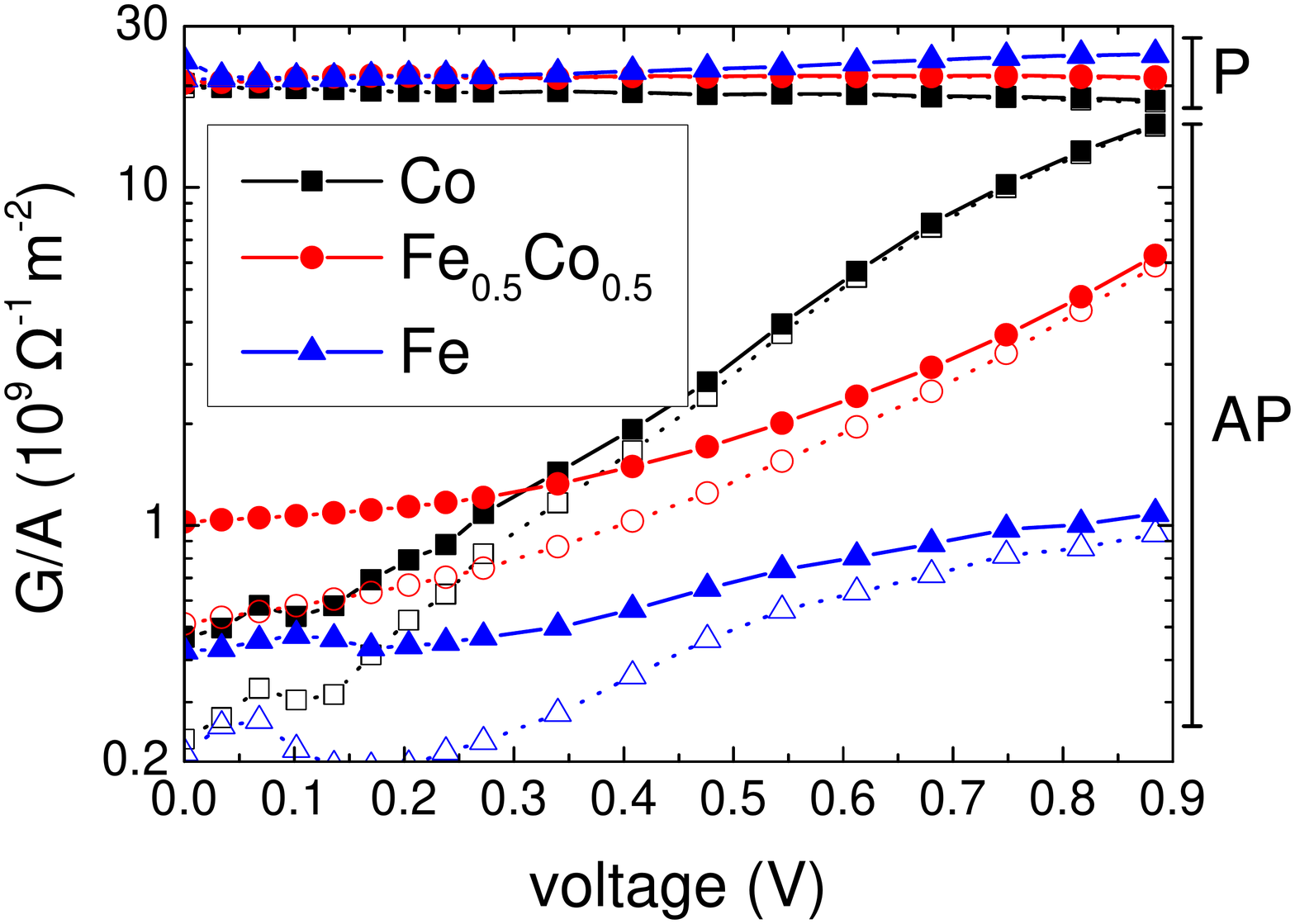}
\caption{(Color online) Bias voltage dependence of the conductances in junctions with Fe, Fe$_{0.5}$Co$_{0.5}$, and Co.
The upper lines are for P and the lower lines for AP magnetizations. The dashed lines indicate the dominating
contribution, which is the $\uparrow\uparrow$ channel for P and the $\downarrow\uparrow$ channel for AP alignment.}
\label{fig_tr_v}
\end{figure}

To understand these very different behaviors, we analyze the bias voltage dependence of the conductances, which enter in
Eq.~(\ref{equ_tmr}). This is shown in Fig.~\ref{fig_tr_v}. The conductance for P alignment is roughly constant for all
three considered junctions. The conductance for AP alignment, on the other hand, shows a very different voltage
dependence for the three materials. For Co leads the AP conductance increases exponentially beyond a certain threshold
voltage and approaches the P value for large voltages. This explains why the TMR drops so drastically for this junction.
As explained above, the AP conductance at zero bias in junctions with Fe$_{0.5}$Co$_{0.5}$ leads is twice as large as for
the pure materials. Nevertheless, the increase with the bias voltage is much slower than for pure Co. Therefore, the Co
AP conductance exceeds the Fe$_{0.5}$Co$_{0.5}$ value at 0.3 V leading to the observed reversal in the TMRs. In comparison,
the increase in the AP conductance for Fe leads by a factor of 2.6 is rather small, inducing a moderate decrease in the
TMR.

\begin{figure}
\includegraphics[width=0.99 \linewidth]{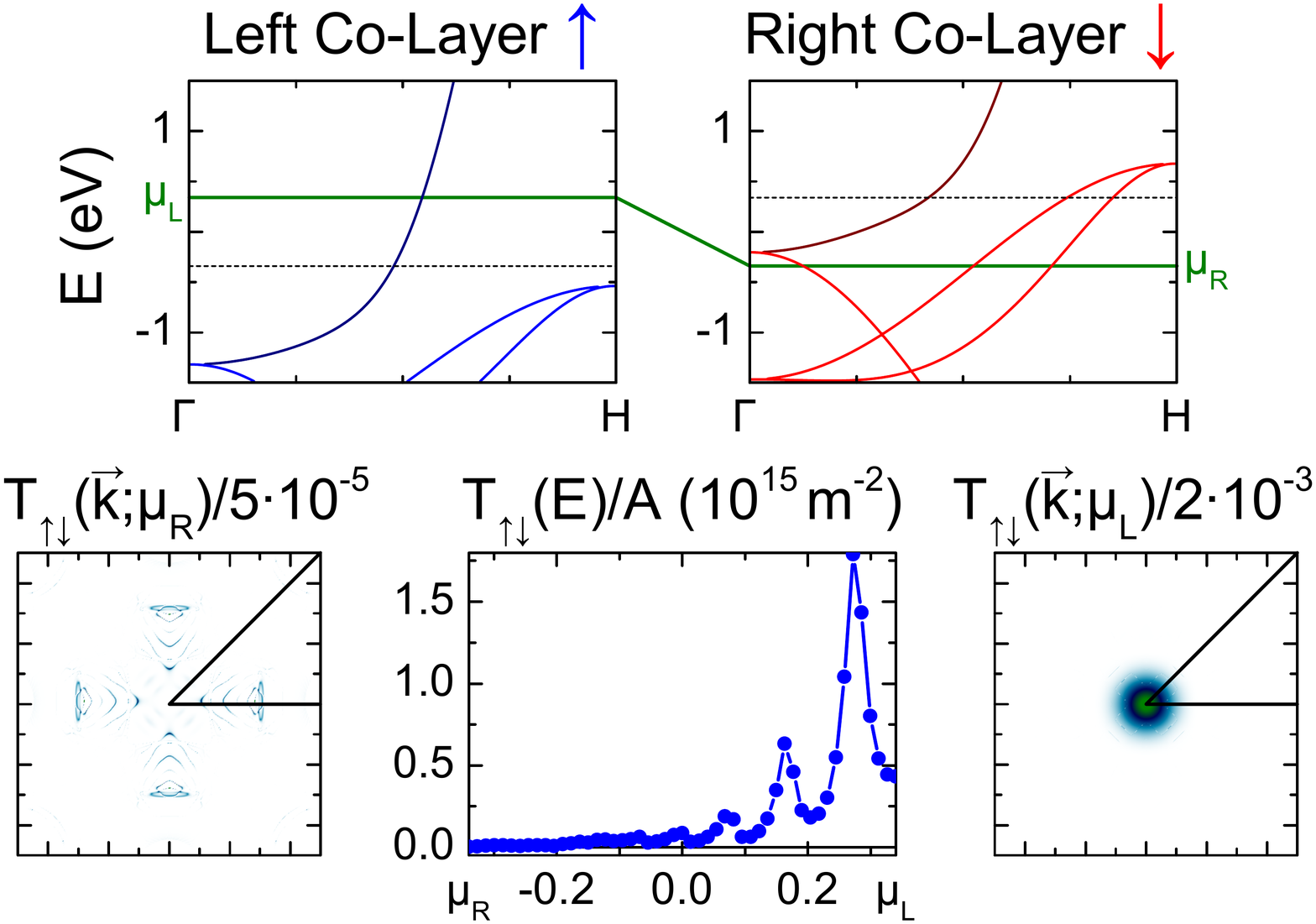}
\caption{(Color online) Top: band alignment for a junction containing 20 monolayers Co (no thickness averaging) with AP
magnetizations in the $\uparrow\downarrow$ channel at a bias voltage of $V=(\mu_R-\mu_L)/e=-0.680\ \mathrm{V}$. The
conduction electrons are moving left to right. Bottom: energy resolved conductance (middle) and
$\vec{k}_\parallel$-resolved conductance at $E=\mu_L$ and $E=\mu_R$ (left and right) in this channel.}
\label{fig_tr_Ex}
\end{figure}

We now explain the reason for the strong increase of the AP current for Co leads. For convenience we consider the case of
a negative bias voltage, i.e.\ the electrons are moving left to right. The increase in $I_{ap}$ in this case is caused by
an increase in the $I_{\uparrow\downarrow}$ channel. As illustrated in Fig.~\ref{fig_tr_Ex}, the applied voltage changes
the band alignment in the two ferromagnetic leads. Furthermore, states from a larger energy range contribute to the
current (Eq.~(\ref{equ_tr})). At zero bias we have $\mu_L=\mu_R=E_F$ thus only states from the $\Delta_1$-half-metallic
region contribute. In particular, majority $\Delta_1$ states cannot contribute for AP alignment, since they are
reflected by the other lead. Increasing the negative bias voltage gradually closes the gap between the majority
$\Delta_1$ states in the left lead and the minority $\Delta_1$ states in the right lead. For Co the gap is closed at a
voltage of $-(E_{\Delta_1 \downarrow}-E_F)/e=-0.134\ \mathrm{V}$, where $E_{\Delta_1 \downarrow}$ is the energy of the
minority $\Delta_1$ band at the $\Gamma$ point. For larger negative voltages an increasing number of $\Delta_1$ states
contributes to the current in the $\uparrow\downarrow$-channel, leading to the observed increase of the AP
current for Co leads. As an example, Fig.~\ref{fig_tr_Ex} shows the band alignment and the contributing energy range
for a large negative voltage. The energy resolved transmission clearly shows the onset of the $\Delta_1$ contribution.
This is superimposed by strong oscillations from quantum well states. Additionally, we show the $\vec{k}_\parallel$
resolved transmission at both endpoints of the energy range. The transmission at $\mu_L$ shows the dominant peak at
$\overline{\Gamma}$ from the $\Delta_1$ states, this is not present at $\mu_R$, which is below the right minority
$\Delta_1$ band. For large positive voltages the same effect occurs in the $I_{\downarrow\uparrow}$ channel.

For Fe the Fermi-energy is in the middle of the $\Delta_1$-half-metallic region and we have $(E_{\Delta_1
\downarrow}-E_F)/e=1.36\ \mathrm{V}$. Thus, the gap is not closed and we do not get large contributions from a
$\Delta_1$-metallic region to the AP current within the considered voltage range.
The Fe$_{0.5}$Co$_{0.5}$ alloy is an intermediate case. Because of the broadening it is difficult to define the bottom of
the $\Delta_1$ band (Fig.~\ref{fig_band}). From the conductance (Fig.~\ref{fig_tr_v}) we find that the onset of the
$\Delta_1$ contributions is delayed compared to Co and much smoother.

The voltage dependence of the TMR for Fe and (bcc) Co leads was investigated experimentally~\cite{yuasa2006}. They
find that the TMR decreases faster with the bias for Co than for Fe, in agreement with our results. However, for Fe the
experiment shows a stronger decrease than our calculation. This indicates additional inelastic effects in particular for
higher voltages. Therefore, in future investigations we plan to include the description of inelastic effects.

The explanation presented in this section is partially in agreement with the one found in Ref.~\onlinecite{bonell2012}.
In addition to the contributions from the $\Delta_1$-metallic region, they propose a strong influence of an interface
resonance state (IRS), which crosses the Fermi-energy from above with increasing Co concentration. In our calculations we
do not see convincing evidence for this effect. We observe the IRS, which is at the Fermi-energy in pure Fe. This IRS
leads to an enhancement in the $\downarrow\downarrow$ channel~\cite{butler2001}. This is visible in
Fig.~\ref{fig_tr_k_all} and \ref{fig_tr_v}. However, we find that this contribution drops quickly with increasing voltage
and Co concentration. Additionally, this IRS also leads to an increase in the AP conductance and both partially cancel
in the TMR. Thus, our calculations indicate that, for the considered barrier thickness of 6 monolayers with perfect
interfaces, the IRSs are of minor importance. However, a dedicated \textit{ab initio} study might be advisable to clarify
the effect of the IRS.

To summarize, we find that different effects control the TMR at zero and large bias. At zero bias the chemical disorder
leads to an increase in the AP conductance and thus a decrease in the TMR for finite concentrations. Even small amounts
of disorder suffice to reduce the TMR to values around 2000\%. At large bias the TMR is controlled by the onset of
contributions from a $\Delta_1$-metallic region to the AP current. The threshold voltage for these contributions is
determined by the distance between the minority $\Delta_1$ band and the Fermi-energy, which decreases with the Co
concentration. Therefore, the TMR decreases with the Co concentration. The main effect of the disorder in this case is to
smoothen the onset. In real junctions one can expect that additional disorder smooths out the peaks in the TMR for small
concentrations and small voltages. In this case, the concentration dependence at small bias would qualitatively follow
the one found at the higher voltage.

\clearpage

\subsection{Spin-Transfer Torque}\label{sec_stt}

\begin{figure}
\includegraphics[width=0.99 \linewidth]{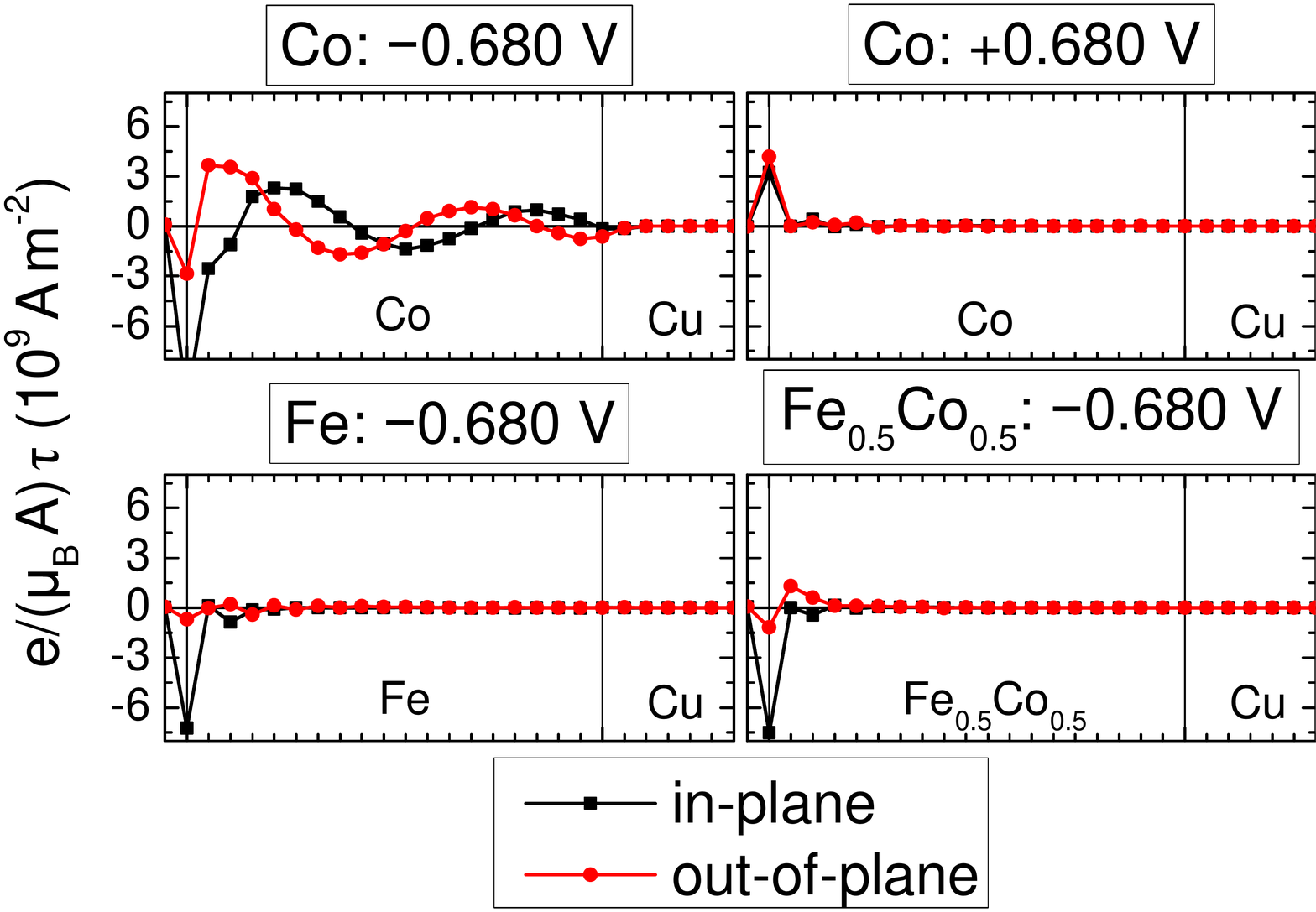}
\caption{(Color online) Layer resolved in-plane and out-of-plane component of the spin-transfer torque in the free layer
for different lead materials and bias voltages}
\label{fig_stt_layer}
\end{figure}

The underlying mechanisms that determine the spin-transfer torque (STT) in FeCo/MgO/FeCo tunnel junctions are closely
related to those responsible for the high TMR. The spin-polarized current through the barrier is dominated by $\Delta_1$
electrons. This leads to a STT which is restricted to the interface~\cite{heiliger2008prl} (see also
Fig.~\ref{fig_stt_layer}). The reason is that the precession of the transport electron is a superposition of the
propagating majority and the evanescent minority state. This leads to a decaying precession so that the torque is
restricted to the interface. In all-metallic systems the restriction to the interface occurs due to
dephasing~\cite{stiles2002,stiles2006}. Dephasing arises from the different precession frequencies of the contributions
from the entire Brillouin zone. However, in tunnel junctions there is only a small number of contributing states and
dephasing is weak. Therefore, the half-metallic nature of FeCo with respect to the dominating $\Delta_1$ states is
important for the STT in such junctions.

\begin{figure}
\includegraphics[width=0.99 \linewidth]{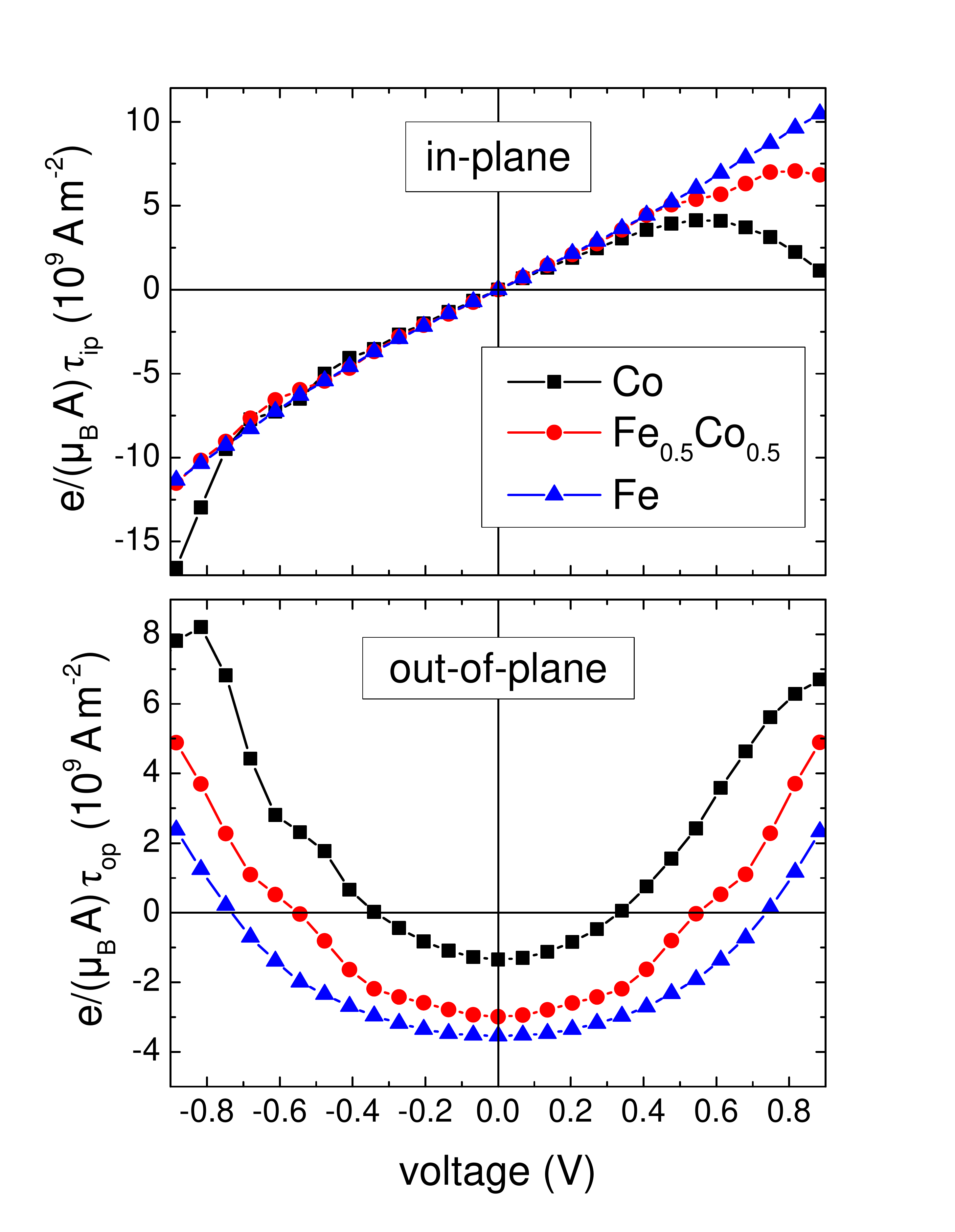}
\caption{(Color online) In-plane and out-of-plane component of the spin-transfer torque as a function of
the applied voltage at a relative angle of $\theta=90^\circ$ between the magnetizations for junctions I-III (see
Fig.~\ref{fig_sketch}), where junction III contains disordered Fe$_{0.5}$Co$_{0.5}$ layers.}
\label{fig_stt_bias}
\end{figure}

Figure~\ref{fig_stt_bias} shows our \textit{ab initio} results for the STT obtained from Eq.~(\ref{eq_torque}) as a
function of the applied bias voltage for junctions I to III. Note that the voltage is going up to $\pm 0.9$ V and
therefore further than in our previous study~\cite{heiliger2008prl} for pure Fe. As in previous studies, we find a simple
bias dependence for pure iron layers. The in-plane torque is almost perfectly linear while the out-of-plane torque is
quadratic. Actually, a convincing fit in the presented voltage range requires a biquadratic polynomial $\tau_{op}(V)
\approx a+b\,V^2+c\,V^4$. For pure cobalt we get a similar behavior for small voltages but strong deviations from the
simple dependence at larger voltages. In particular, for the in-plane STT we find a strong reduction for large positive
bias and an enhancement for large negative bias. These deviations are the result of contributions from the
$\Delta_1$-metallic regime to the AP current (compare Sec.~\ref{sec_tmr}). The latter lead to a large and highly
spin-polarized AP current, which cancels (adds up) with the spin-polarized P current for positive (negative) bias.

Figure~\ref{fig_stt_bias} includes the bias dependence for junction III with disordered Fe$_{0.5}$Co$_{0.5}$ layers. We
find almost the simple behavior of pure iron with only small deviations at larger voltages, which are weaker and smoother
compared to pure cobalt. In this computationally very demanding calculation, we omit the thickness averaging. The
disorder in Fe$_{0.5}$Co$_{0.5}$ reduces the quantum well effects, only small residual oscillations are visible. The
smooth bias dependence can be explained directly by looking at the Bloch spectral density of Fe$_{0.5}$Co$_{0.5}$, which
is shown in Fig.~\ref{fig_band}. It shows a strong broadening for some of the bands caused by the disorder scattering. In
particular the minority $\Delta_1$ band shows a strong broadening. This leads to a smooth transition from the
$\Delta_1$-half-metallic to the $\Delta_1$-metallic regime and thus explains the smooth onset of the deviations (as
explained in Sec.~\ref{sec_tmr}). Approximating the alloys with ``ordered alloys" leads to larger and more complicated
deviations (not shown). Therefore, an accurate description of the alloy scattering is necessary to obtain the correct
voltage dependence.

We find that the STT is of the same order of magnitude for all junctions under investigation. The observed voltage
dependence can be qualitatively explained from the band structure or the Bloch spectral density. The bias dependence of
the STT using Fe$_{0.5}$Co$_{0.5}$ layers is quite similar to that for pure iron layers. Thus, our results can explain the
agreement between our previous \textit{ab initio} results~\cite{heiliger2008prl} and the experiment~\cite{sankey2008}
although both investigations use different ferromagnetic materials.

\begin{figure}
\includegraphics[width=0.99 \linewidth]{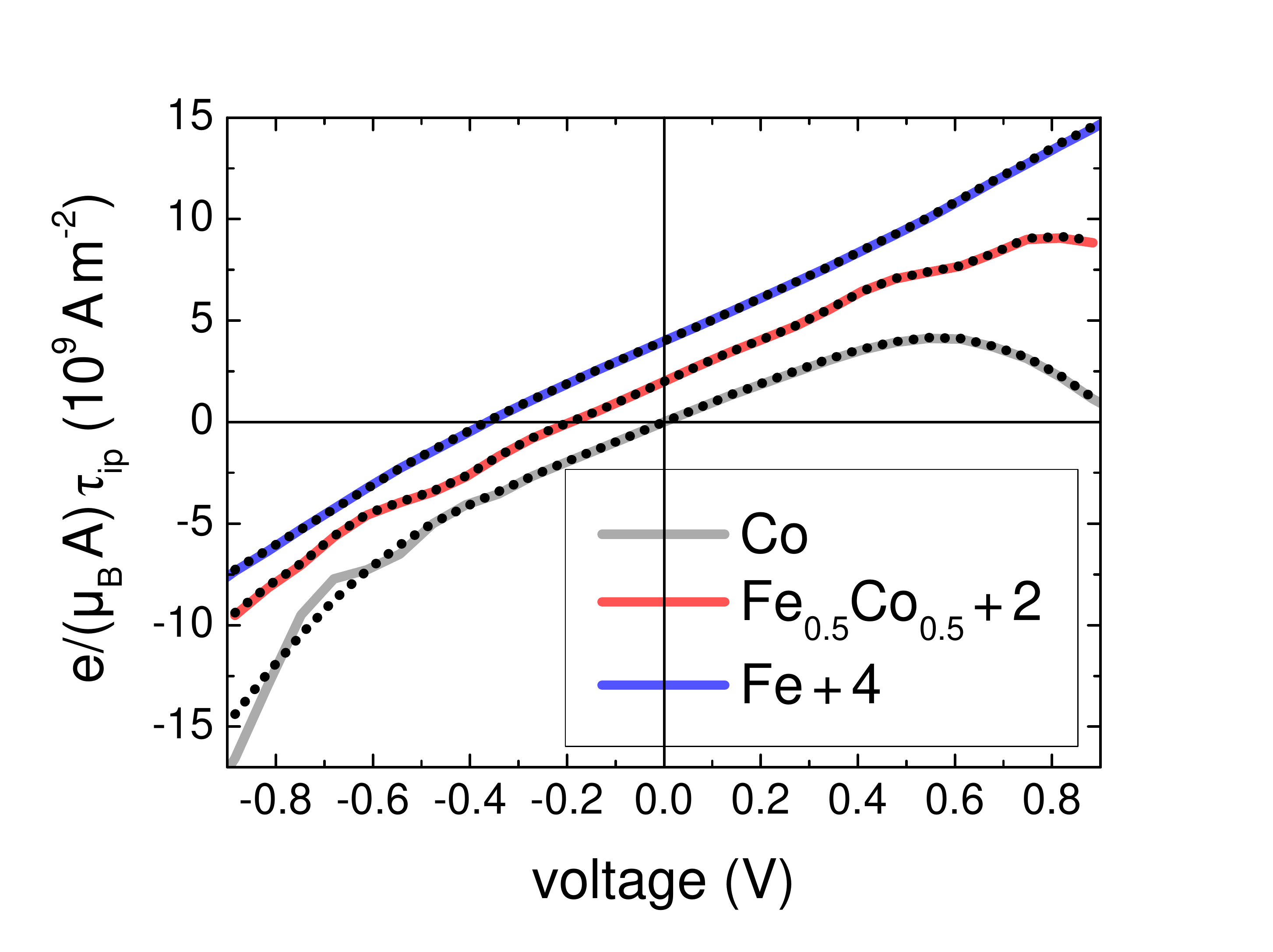}
\caption{(Color online) Comparison of the in-plane component of the spin-transfer torque calculated from the
non-equilibrium density using Eq.~(\ref{eq_torque}) as in Fig.~\ref{fig_stt_bias} (solid) and from the spin currents
using Eq.~(\ref{Is_to_STT}) (dotted). The curves are shifted to improve visibility.}
\label{fig_STT_Is}
\end{figure}

\begin{figure}
\includegraphics[width=0.99 \linewidth]{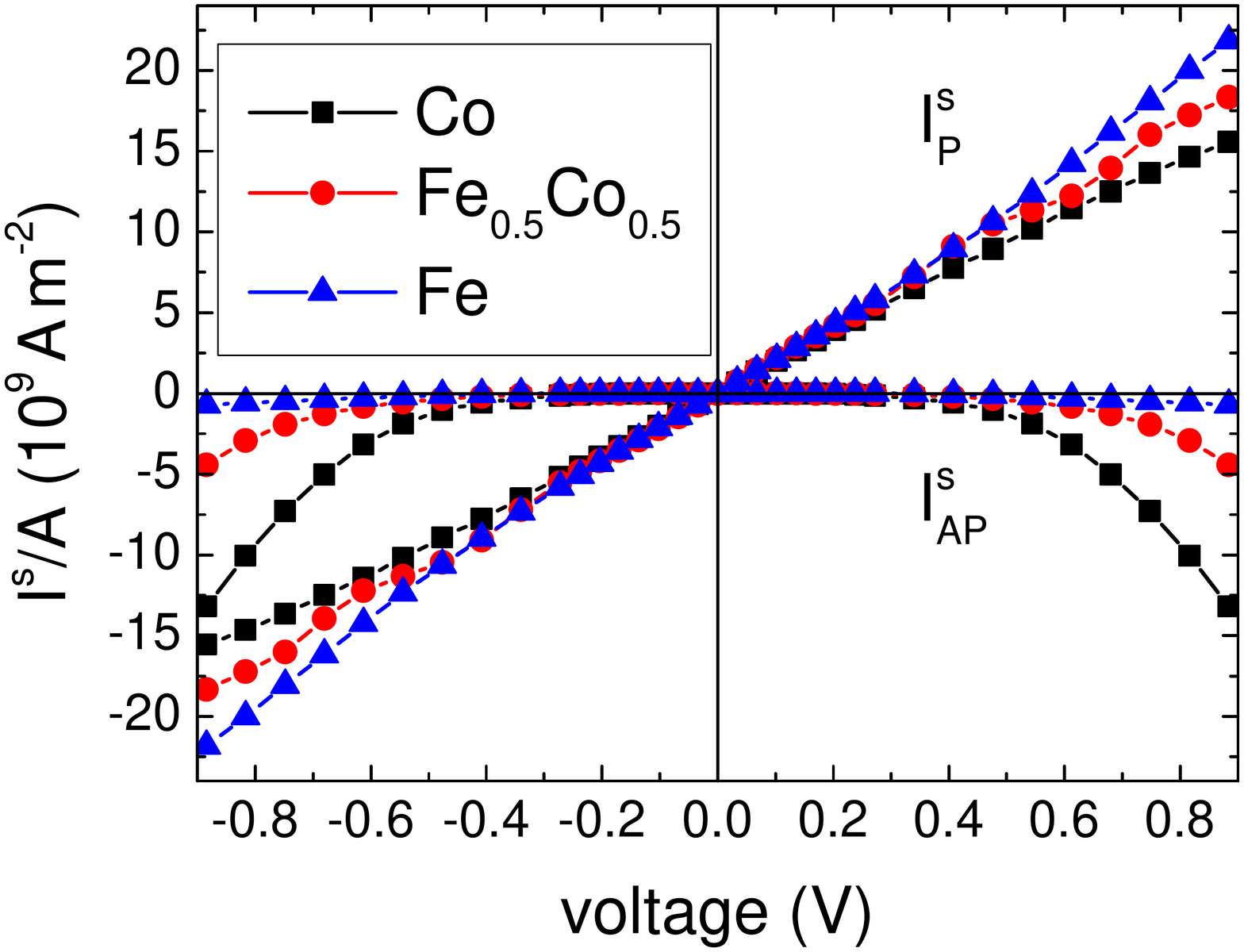}
\caption{(Color online) Spin currents through the junctions for P and AP alignment.}
\label{fig_Is}
\end{figure}

To provide a quantitative explanation, the in-plane component of the STT can be described by a simple expression in terms
of spin currents~\cite{slonczewski2005,theodonis2006}. If the current in the left (right) ferromagnetic lead is
determined far enough from the barrier, its polarization will be aligned with the local magnetization and we can define
the spin current as $I^s_{L(R)}=I^\uparrow_{L(R)}-I^\downarrow_{L(R)}$. By conservation of angular momentum, the
difference in the spin currents in right and left lead has to be absorbed by the magnetizations and thus creates the STT.
This leads to the expression~\cite{slonczewski2005}
\begin{equation} \label{Is_to_STT_000}
\tau_{ip}(\theta)=\frac{\mu_B}{e} \frac{1}{\sin(\theta)} \left(I^s_L(\theta)-I^s_R(\theta) \cos(\theta) \right).
\end{equation}
The spin currents depend on the relative angle $\theta$ between the magnetizations.
Making use of the general transformation of a spin state under a rotation, this expression can be simplified to a form
that only uses the spin currents in the P and AP alignment~\cite{slonczewski2005}
\begin{equation} \label{Is_to_STT}
\tau_{ip}(\theta)=\frac{1}{2}\frac{\mu_B}{e}\left(I^s_P + I^s_{AP}\right) \sin(\theta),
\end{equation}
where the spin currents can be determined from the four spin channels introduced in Sec.~\ref{sec_method}:
$I^s_{P}=I^{\uparrow\uparrow}-I^{\downarrow\downarrow}$ and $I^s_{AP}=I^{\uparrow\downarrow}-I^{\downarrow\uparrow}$. The
in-plane component of the STT calculated from the spin currents is compared to the results from Eq.~(\ref{eq_torque}) in
Fig.~\ref{fig_STT_Is}. We find perfect agreement, except for Co at large negative bias. For this case the contributions
to the STT do not completely decay inside the ferromagnetic layer. This can be observed in Fig.~\ref{fig_stt_layer},
which shows the layer-resolved torque for different cases. Thus, the prerequisites of Eq.~(\ref{Is_to_STT_000}) are not
strictly fulfilled. The description in terms of spin currents provides a quantitative explanation of the effects that
determine the in-plane STT. The spin currents entering in Eq.~(\ref{Is_to_STT}) are calculated from the data obtained for
the TMR in Sec.~\ref{sec_tmr} and are shown in Fig.~\ref{fig_Is}. While the spin currents for the P alignment are roughly
linear for AP they show a nonlinear increase in negative value, which is strongly enhanced from Fe to
Fe$_{0.5}$Co$_{0.5}$ to Co. This is caused by the increase in the $\downarrow\uparrow$ ($\uparrow\downarrow$) channel for
positive (negative) bias which, as explained in Sec.~\ref{sec_tmr}, is due to contributions from a $\Delta_1$-metallic
regime. This explains the attenuation of the in-plane STT for positive voltages and the enhancement for negative voltage,
which is most pronounced for Co. From the derivation and the persuasive agreement in Fig.~\ref{fig_STT_Is} we can assume
that the validity of Eq.~(\ref{Is_to_STT}) will hold for all angles and thus an investigation of the angular dependence
is omitted.

The same arguments that lead to Eq.~(\ref{Is_to_STT_000}) also yield an identity for the out-of-plane
component~\cite{slonczewski2005}
\begin{equation} \label{eq_STT_op}
\tau_{op,R}=-\tau_{op,L},
\end{equation}
which relates the torques exerted on both ferromagnetic layers. Note that this does not require a symmetric junction.
This equation is fulfilled accurately for junctions containing Fe and Fe$_{0.5}$Co$_{0.5}$ leads, but only for low
voltages in junctions with Co leads, because for higher voltages the contributions to the torque do not fully decay
inside the ferromagnetic lead (Fig.~\ref{fig_stt_layer}).

\begin{figure}
\includegraphics[width=0.99 \linewidth]{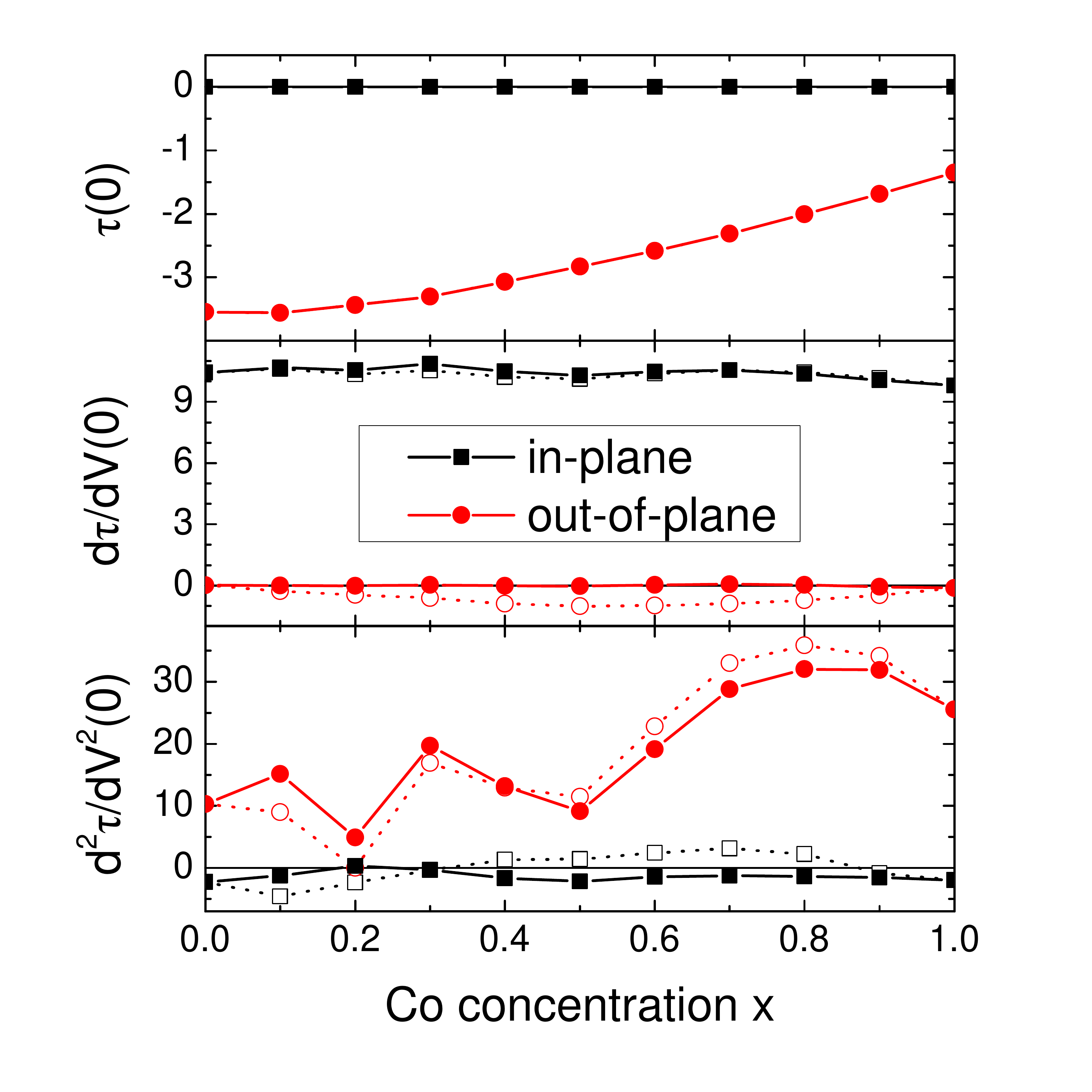}
\caption{(Color online) Concentration dependent coefficients in an expansion of the spin-transfer torque about zero
bias: $\tau(V) \approx \tau(0) + \frac{\mathrm{d}\tau}{\mathrm{d} V}(0)\,V + \frac{1}{2}\frac{\mathrm{d^2}\tau}{\mathrm{d} V^2}(0)\,V^2$. 
The shown data is scaled with a prefactor $e/(\mu_B \, A)$ and has units of 
$10^{9}\,\mathrm{A}\,\mathrm{m}^{-2}$, 
$10^{9}\,\mathrm{\Omega}^{-1}\,\mathrm{m}^{-2}$, and
$10^{9}\,\mathrm{\Omega}^{-1}\,\mathrm{V}^{-1}\,\mathrm{m}^{-2}$.
The dashed lines indicate the results neglecting diffusive contributions.}
\label{fig_STT_c}
\end{figure}

To gain insight into the concentration dependence of the STT we calculate an expansion about zero bias for the full
range. This is obtained from a quadratic fit in a small voltage range ($\pm 68.0 \ \mathrm{mV}$) and shown in
Fig.~\ref{fig_STT_c}. As expected, the in-plane component is zero for all concentrations. We find that the out-of-plane
component (i.e.\ the interlayer exchange coupling) decreases in negative value with the Co concentration. The first
derivative of the out-of-plane component is zero by symmetry and for the in-plane component it is constant. This has
already been noted above and has important implications for optimizing devices. The second derivative determines the
quadratic component and thus it is almost zero for the in-plane component. For the out-of-plane component it has the same
order for all concentrations and shows a slight increase with the Co concentration.

When we compare these results with the concentration dependence of the TMR at zero bias (Fig.~\ref{fig_TMR_c}), the
absence of any large changes between the pure limits and finite concentrations is conspicuous. As shown before, the
concentration dependence of the TMR is mostly determined by the AP-conductance, which in turn shows a strong increase
from zero to finite concentrations (Fig.~\ref{fig_trC_V1}). However, the STT for small voltages is completely dominated
by the $\Delta_1$ states which determine the P-conductance and are only weakly affected by the alloy concentration. This
can also be seen from Eq.~(\ref{Is_to_STT}): The P spin current $I^s_P$ is dominated by the $\Delta_1$ states, while
$I^s_{AP}$ vanishes at zero bias.

Figure~\ref{fig_STT_c} also shows the results obtained without diffusive contributions, i.e.\ neglecting the NVC. The
deviations seem rather small, but from the nonvanishing first derivative of the out-of-plane component we infer that this
approximation leads to a systematic error and even to a violation of the symmetry $\tau_{op}(-V)=\tau_{op}(+V)
\Rightarrow \frac{\mathrm{d}\tau_{op}}{\mathrm{d} V}(0)=0$, which follows from Eq.~(\ref{eq_STT_op}) for symmetric
junctions.

\section{Conclusion}
Our calculations for the TMR at zero bias show very large values for pure Fe and pure Co leads, which were
previously reported in literature. However, even small amounts of chemical disorder caused by alloying lead to a large drop
resulting in a TMR of about 2000\% for all finite concentrations. This drop is a consequence of the disorder
scattering, which leads to a redistribution of the states in $\vec{k}_\parallel$-space and to an increased overlap of the
states in the antiparallel alignment. Since small amounts of disorder are hard to avoid in real junctions, this
calculated value might pose a more realistic limit for what can be achieved. Nevertheless, it is still a factor of two
larger than current experimental record values.

At a large bias voltage, we find a decrease of the TMR with the Co concentration. This is caused by minority $\Delta_1$
states, which enter the energy window for transport at high Co concentration and finite bias voltage. This contribution
is an inevitable consequence of the band filling and thus the optimum TMR should be found at small to zero Co
concentration. This is in contradiction to experimental results~\cite{lee2007,bonell2012}, which find a maximum TMR
for about 25\% Co. In our calculations we assume ideal interfaces. Therefore, a possible explanation of this
discrepancy is that the quality of the real interfaces depends on the concentration. In this case, a detailed
investigation of the concentration dependence of the quality of the interfaces should clarify the discrepancy.

The in-plane (out-of-plane) component of the STT shows the expected linear (quadratic) bias dependence at small voltages.
At large voltages and large Co concentrations we find a strong deviation from this simple dependence. By using an
expression in terms of the spin-currents in the P and AP alignment, this is traced back to the same effects, which govern
the TMR at large voltages. Since the STT at small bias turns out to be mostly independent of the composition the
optimization can be focused on the TMR as long as switching can be achieved below the onset of the nonlinear deviations in the voltage
dependence.

We find that in all calculations the diffusive contributions (vertex corrections) are important. While for the TMR
neglecting them leads to meaningless results, for the STT it leads to relatively small errors, which however beak
physical symmetries.

\section{Acknowledgement}
We acknowledge support from DFG grant \mbox{HE 5922/1-1}.

\bibliography{bib}

\end{document}